\documentclass[]{svjour2}
\smartqed
\usepackage{graphicx}% Include figure files
\usepackage{dcolumn}% Align table columns on decimal point
\usepackage{color}% color package
\usepackage{hyperref}% hyper link cite package
\usepackage{latexsym}
\usepackage{amsmath}
\journalname{}
\begin{document}

\title{ Steady state of Stochastic Sandpile  Models}

\author{Tridib Sadhu \and 
Deepak Dhar
}
\institute{Tridib Sadhu \at
Department of Theoretical Physics, \\  
Tata Institute of Fundamental Research, \\
Homi Bhaba Road, Mumbai 400005, India. \\
\email{tridib@tifr.res.in} 
\and
Deepak Dhar \at
Department of Theoretical Physics, \\  
Tata Institute of Fundamental Research, \\
Homi Bhaba Road, Mumbai 400005, India. \\
\email{ddhar@theory.tifr.res.in}
}

\date{\today}

\maketitle

\begin{abstract}
We study the steady state of the abelian sandpile models with stochastic 
toppling rules. The particle addition operators commute with each other, 
but in general these operators need not be diagonalizable. We 
use their abelian algebra to determine their eigenvalues, and the 
Jordan block structure. These are then used to determine the
probability of different configurations in the steady state. We
illustrate this procedure by explicitly determining the numerically
exact steady state for a one dimensional example, for systems of
size $\le12$, and also study the density profile in the steady state.
\end{abstract}

\section{Introduction}
Sandpile models with stochastic toppling rules are important subclass of 
sandpile models \cite{deepak}.  The first such model was studied by 
Manna \cite{manna}, and  these are usually known as Manna models in
the literature. They are able to describe the avalanche 
behavior seen experimentally in the piles of granular media much better than 
the deterministic models \cite{rice}. Also, in numerical studies, one 
gets better scaling collapse, and consequently, more reliable estimates 
for the values of the critical exponents, than for models with deterministic 
toppling rules \cite{chessa,campelo}.

Unfortunately, at present, the theoretical understanding of models with 
stochastic toppling rules is much less than that of their deterministic 
counterparts, e.g. the Bak-Tang -Wiesenfeld (BTW) model \cite{btw}.  For 
example, there is no analogue of the burning test  
to distinguish the transient and  the recurrent states of a 
general Manna model. For the deterministic case, it is known that 
all the recurrent configurations occur with equal probability in the 
steady state. A similar characterization of the steady state is not 
known in the Manna case.  The steady state has been 
explicitly determined only for the fully directed stochastic models 
\cite{asap,rittenberg}. In some cases, one can formally characterize the 
recurrent states of the model, e.g. the $1$-dimensional Oslo rice pile 
model, but a straightforward direct depth-first
calculation of the exact
probabilities of different configurations in the steady state takes 
$\mathcal{O}(exp{(L^3)})$ steps where $L$ is the system length \cite{oslo}. While 
the exact values of the critical exponents have been conjectured for $(1+1)$ 
dimensional directed Manna model \cite{kloster,paczuski}, the 
prototypical undirected Manna model in one dimension has resisted an 
exact solution so far \cite{dickman1,dickman2,dickman3}. In higher 
dimensions, most of the studies are only numerical, and deal with the 
estimation of the critical exponents of avalanche distribution, and the 
universality class of the model 
\cite{chessa,pietronero,biham,lubeck,menech,mohanty1,bonachela,mohanty2}.

While the original Manna model did not have the abelian property of
the BTW model, one can construct stochastic toppling rules with
abelian property \cite{sasm}. In this paper, we discuss this abelian
version of the stochastic Manna model. It is a
special case of the more general Abelian
Distributed Processors Model (ADP) \cite{deepak}. We 
shall use the terms Deterministic Abelian Sandpile Models (DASM) and 
Stochastic Abelian Sandpile Models (SASM), if we need to distinguish 
between these two classes of models.

We use the algebra of the addition operators to determine the steady 
state of the model. This algebraic approach provides a computationally 
efficient method to determine the Markov evolution matrix of the model. 
The addition operators of SASM are not necessarily diagonalizable even 
if we restrict ourselves to the space of recurrent configurations. Using 
the abelian algebra we determine a generalized eigenvector basis in 
which the operators reduce to Jordan block form. We also define a 
transformation matrix between this basis and the configuration basis, 
and express the steady state in the latter basis. This procedure is 
illustrated by explicitly writing out the case of a one dimensional 
Manna model. In this special case, we can show that each Jordan block is 
at most of dimension $2$. We determine the numerically exact steady 
state of the model for systems of size up to $12$ and determine the 
asymptotic density profile by extrapolating the results.

This paper is organized as follows: In section $2$, we define the model 
precisely. In section $3$, we define the addition operators for the 
model, and discuss their algebra. Calculation of the eigenvalues and the 
Jordan block structure of the addition operators are given in section 
$4$. The transformation matrix between the generalized eigenvector basis 
and the configuration basis is determined in section $5$ and is used to 
determine the steady state vector in the configuration basis in section 
$6$. The exact numerical determination of the steady state is discussed 
in section $7$ with some concluding remarks in section $8$.
 
\section{The Model}
 
 We define a generalized Manna model on a graph of $N$ sites with a 
non-negative integer height variable $z_i$ defined at each site $i$. Let 
the threshold height at $i$ be $z_i^c$, and the site is unstable if $z_i 
\geq z_i^c$.  If the system is stable, a sand grain is added at a 
randomly chosen site which increases the height by $1$.  For each site 
$i$, there is a set of $\alpha_i^{max}$ lists $E_{\alpha,i}$ with 
$\alpha=1,2,\cdots,\alpha_i^{max}$.  If a site is unstable, it relaxes 
by the following toppling rule: we decrease its height by $z_{i}^c$. 
Then, with probability $p_{\alpha,i}$, we select the list 
$E_{\alpha,i}$, independent of any previous selections, and then add one 
grain to each site in that list. If a site occurs more than once in the 
list, we add that many grains to that site.

 Toppling at a site can make other sites unstable and they topple in 
their turn, until all the lattice sites are stable.  It follows from the 
abelian property of the model that the probabilities of different final 
stable configurations are independent of the order in which different 
unstable sites are toppled.

We illustrate these rules with some examples below.
\begin{itemize}
\item
\textbf{Model A} (The one dimensional Manna model): The graph is $L$
sites on a line and $z_i^c= 2$, for all sites. On toppling each
grain is transfered to its neighbors with equal probability. Hence
we have $\alpha_i^{max}=3$, for all $i$, with $E_{1,i}=\{i-1, i-1\}$, 
$E_{2, i}=\{i-1,
i+1\}$, and $E_{3, i}=\{i+1, i+1\}$ and
$p_{1,i}=p_{3,i}=1/4$ and $p_{2,i}=1/2$. Also grains can move out of the
system  if  toppling occurs  at a boundary site.
\item
\textbf{Model B} (The one dimensional dissipative Manna model): Same
as model A except that on toppling a grain
can move out of the system with probability $\epsilon$. Then
$\alpha_i^{max}=6$ and the lists of neighbors $E_1=\{i-1,i-1\}$,
$E_2=\{i-1,i+1\}$, $E_3=\{i+1,i+1\}$, $E_4=\{i-1\}$, $E_5=\{i+1\}$ and
$E_6=\Phi$, where $\Phi$ is an empty set. The corresponding probabilities are
$p_{1,i}=p_{3,i}=(1-\epsilon)^2/4$, $p_{2,i}=(1-\epsilon)^2/2$,
$p_{4,i}=p_{5,i}=\epsilon(1-\epsilon)/2$ and $p_{6,i}=\epsilon^2$.

In this case, one can use periodic boundary conditions, as there is 
dissipation at all sites. the steady state is critical only in the limit 
$\epsilon \rightarrow 0$.
For the models A and B, it is easy to see that all stable configurations 
occur in the steady state with non-zero probability. We can also define 
stochastic models where the recurrent configurations form only an 
exponentially small fraction of all stable configurations. An example of 
this type is 

\item
\textbf{Model C}:
The graph is a square lattice with $N$ sites and
$z_{\mathbf{i}}^c=2$. Under
toppling, with equal probability two particles are transfered to
either horizontal or vertical neighbors.  Hence $\alpha_i^{max}=2$ with
$E_{1,\mathbf{i}}=\{\mathbf{i}+\mathbf{e}_x,\mathbf{i}-\mathbf{e}_x\}$
and $E_{2,\mathbf{i}}=\{\mathbf{i}+\mathbf{e}_y,\mathbf{i}-\mathbf{e}_y\}$ with
$p_{1,\mathbf{i}}=p_{2,\mathbf{i}}=1/2$.
\end{itemize}

In the following we will mostly confine ourselves to Model A. The
treatment of other cases presents no special difficulties.

\section{The addition operators and their algebra}
Let us denote the space of stable states as $\Gamma$ spanned by
$\Omega=\displaystyle \prod_{i=1}^{N}z_i^c$
basis vectors labeled by $C$. We define $P(C,t)$ as the probability of
finding the system in the basis $C$ at time $t$. To each set
$\{P(C,t)\}$, we associate a vector $|P(t)\rangle$ belonging to the
vector space $\Gamma$, and write 
\begin{equation}
  |P(t)\rangle = \sum_C P(C,t)|C\rangle.
  \label{P(t)}
\end{equation}
We define the particle addition operators $\textbf{a}_i$ for all $i$ 
as linear operators acting on $\Gamma$ as follows:
Consider adding a sand grain at site $i$
in a configuration $C$, and relaxing the system until a stable
configuration is reached. For stochastic toppling rules, the 
resulting state is not necessarily a basis vector corresponding
to a unique stable configuration, but a linear combination of
them. If the resulting configuration is $C'$
with probability $P_i(C'|C)$, we define
\begin{equation}
  \mathbf{a}_i|C\rangle = \sum_{C'} P_i(C'|C) |C'\rangle,
  \label{a_i}
\end{equation}
for all $C$. Note that the action of any of these operators on a given
configuration gives a unique probability state vector. 

Eq. (\ref{a_i}) is a formal definition of the operators 
$\{\mathbf{a}_i\}$. One can think of these as $\Omega \times \Omega$ 
matrices, but  it is quite non-trivial to actually determine
the matrix elements $P_i(C'|C)$ explicitly from the toppling rules. This 
is 
because of the non-zero probability of an arbitrary large 
number of toppling before a steady state is reached.

For an example, consider the avalanches in model A for system of size $L=3$.
Consider the $2^3$ stable configurations as the basis vectors 
and denote them by their height values $|z_1, z_2, z_3 \rangle$.
The action of $\textbf{a}_2$ on $|0,1,0\rangle$ will
generate a unstable state $|0,2,0\rangle$. Using the toppling rules
we can write the following set of equations for three unstable states
\begin{eqnarray}
  |0, 2, 0\rangle &=& \frac{1}{4}|2, 0, 0\rangle + \frac{1}{2}|1, 0,
1\rangle + \frac{1}{4}|0, 0, 2\rangle, \nonumber \\
  |2, 0, 0\rangle &=& \frac{1}{4}|0, 2, 0\rangle + \frac{1}{2}|0, 1,
0\rangle + \frac{1}{4}|0, 0, 0\rangle, \nonumber \\
  |0, 0, 2\rangle &=& \frac{1}{4}|0, 2, 0\rangle + \frac{1}{2}|0, 1,
0\rangle + \frac{1}{4}|0, 0, 0\rangle.
  \label{|0,2,0}
\end{eqnarray}
We see that there is a nonzero probability that the avalanche
can continue for more than $s$ toppling, for any finite $s$. e.g.  in 
the sequence 
$|0,2,0 \rangle \rightarrow |2,0,0\rangle \rightarrow |0,2,0\rangle \cdots$. 
Thus straight forward application of the relaxation rules do not result 
in a finite procedure to determine the unstable vector $|0,2,0\rangle$
in terms of the stable configurations. Instead, we have to  write Eq. 
(\ref{|0,2,0})
as a matrix equation
\begin{align}
 \textbf{M} 
  \begin{bmatrix}
    |0,2,0\rangle \\
    |2,0,0\rangle \\
    |0,0,2\rangle
 \end{bmatrix}
  =
  \begin{bmatrix}
    |1,0,1\rangle \\
    |0,1,0\rangle \\
    |0,0,0\rangle
 \end{bmatrix},
\end{align}
and then invert it.
More generally, the determination of $P(C'|C)$ involves working in
a large space of unstable configurations. 

For example in model A, there are $2^L$  stable configurations, 
where each site has $0$ or $1$ particle. Total number of particles is at 
most $L$. On adding one particle, the  number of 
particles can become $L+1$, where initially, only one site will have 
height $2$. However, it is easy to verify that by toppling one can 
generate configurations  where the number of particles at a site is much 
greater than $2$. In fact,  all the $L+1$ particles could be at the same 
site. Then the total number of stable and unstable configurations  
$\Omega'$ is the number of 
ways one can distribute $L+1$ particles on $L$ sites . It is easily 
seen that $\Omega'$ varies as $4^L$, and one
needs to invert a matrix of size $\mathcal{O}(\Omega' \times \Omega')$.

In this paper we will use the operator algebra
to obtain an efficient method to determine the probabilities $P(C'|C)$
explicitly which requires inverting a matrix only of size $2^L \times 2^L$.
It has been shown \cite{sasm} that the addition operators for different sites
commute i.e.
\begin{equation}
  [\mathbf{a}_i, \mathbf{a}_j] =0 \rm{,~ ~ ~ for ~ all ~} i,j.
\end{equation}
Unlike the DASM, the inverse operators $\{\mathbf{a}^{-1}_i\}$
for SASM need not exist, even if we restrict ourselves to the set of 
recurrent configurations. This is because among the recurrent
states, one can have two different
initial probability vectors that yield the same resultant vector.
This makes the determination of the matrix form of the operators
difficult for this model.

Apart from the abelian property, the operators also satisfy a set 
of algebraic equations. For simplicity of presentation, now on we
consider $z_i^c=z_c$ and $p_{\alpha,i}=p_{\alpha}$ for all sites. Then
consecutive addition of $z_c$ grains at a site 
ensures that the site will topple once and transfers $z_c$ grains to
its neighbors, irrespective of the initial height. Then
the operators obey the following equation
\begin{equation}
  \mathbf{a}_i^{z_c} =
\sum_{\alpha}p_{\alpha}\mathbf{a}^{E_{\alpha, i}} \rm {~for~
}1\le i\le N, 
\end{equation}
where we have used the notation $\mathbf{a}^{E}=\displaystyle
\prod_{x\epsilon E}\mathbf{a}_{x}$ for any list $E$, and
\begin{equation}
  \mathbf{a}_i = \mathbf{1},
  \label{a_0}
\end{equation}
for sites $i$ outside the lattice. In particular for the examples
in Section $2$, these equations are as follows
\begin{eqnarray}
\mathbf{a}_i^2&=&\frac{1}{4}(\mathbf{a}_{i-1}+\mathbf{a}_{i+1})^2
\rm{~~~~~~~~~~~~~~~~~~~~~~~for~Model~A,} \label{modA} \\
\mathbf{a}_i^2&=&[\frac{1-\epsilon}{2}\mathbf{a}_{i-1}+\frac{1-\epsilon}{2}\mathbf{a}_{i+1}+\epsilon
\mathbf{1}]^2 \rm{~~~~~for~Model~B,~and~} \\
\mathbf{a}_\mathbf{i}^2&=&\frac{1}{2}(\mathbf{a}_{\mathbf{i}-\mathbf{e}_x}\mathbf{a}_{\mathbf{i}+\mathbf{e}_x}+\mathbf{a}_{\mathbf{i}-\mathbf{e}_y}\mathbf{a}_{\mathbf{i}+\mathbf{e}_y})\rm{~~~~~~~~for~Model~C.}
\end{eqnarray}

\section{Jordan Block structure of the addition operators}

In general the matrices $\{\mathbf{a}_i\}$ need not be diagonalizable. 
However, using the abelian property, we can construct a common set of
generalized eigenvectors for all the operators $\{a_i\}$ such that in this basis
the matrices simultaneously reduce to Jordan block form. These 
generalized eigenvectors split the vector space $\Gamma$ into
disjoint subspaces, each corresponding to distinct set of eigenvalues.
There will be at least one common eigenvector in
each subspace, for all the addition operators. 

Proof: Consider one of the operators, say $\mathbf{a}_1$.
Let $\Gamma_1$
be the subspace of $\Gamma$ spanned by the (right) generalized eigen vectors of 
$\mathbf{a}_1$ corresponding to the eigenvalue $a_1$. There is at
least one such generalized eigenvector, so $\Gamma_1$ is non-null. We pick one
of the other addition operators, say $\mathbf{a}_2$. From the fact that
$\mathbf{a}_2$ commutes with $\mathbf{a}_1$, it immediately follows
that $\mathbf{a}_2$ acting on any vector in the subspace $\Gamma_1$
leaves it within the same subspace. Diagonalizing $\mathbf{a}_2$
within this subspace, we construct a possibly smaller but still
non-null subspace $\Gamma_2$ which is spanned by simultaneous 
eigenvectors of $\mathbf{a}_1$ and $\mathbf{a}_2$ with eigenvalues
$a_1$ and $a_2$. Repeating this argument with the other operators,
one can construct vectors which are simultaneous eigenvectors of 
all the $\{\mathbf{a}_i\}$. 

Let $|\psi \rangle$ be such an eigenvector, with
\begin{equation}
  \mathbf{a}_i|\psi\rangle = a_i |\psi \rangle \rm{,~ ~ ~ for ~} \rm{~
}\rm{~}1 \le i \le N. 
\end{equation}
Then from Eq.($6$)the eigenvalues satisfy the following set of equations
\begin{equation}
  a_i^{z_c} = \sum_{\alpha}p_{\alpha}a^{E_{\alpha,i}}\rm{~ ~ ~ for
~ } \rm{~ }\rm{~}1 \le i \le N,
\end{equation}
where we have used the notation
$a^{E}=\displaystyle\prod_{x\epsilon E}a_{x}$, for any list $E$.

 Rather than work with this general case, we will consider the special
case in model A for simplicity. No extra complications occur in
the more general case. Then, from Eq.(\ref{modA}), the corresponding
eigenvalue equation is
\begin{equation}
  a_i^2 = \frac{1}{4}(a_{i-1}+a_{i+1})^2 \rm{,~ ~ ~ for ~ } \rm{~ }\rm{~}1 \le i \le L 
  \label{eveqn}
\end{equation}
These are L coupled quadratic equations in $L$ complex variables $\{a_i\}$.
We can reduce them to $L$ linear equations by taking square root
\begin{equation}
  \eta_ia_i=\frac{1}{2}(a_{i-1}+a_{i+1}),
  \label{eta}
\end{equation}
where $\eta_i=\pm1$. The Eq. (\ref{a_0}) sets the values for the 
eigenvalues of $\mathbf{a}_0$ and $\mathbf{a}_{L+1}$ which are 
\begin{equation}
  a_1=a_{L+1}=1.
  \label{a_1}
\end{equation}
There are $2^L$ different choices for the set of $L$
different $\eta$'s and for each such choice, we get a set of
eigenvalues $\{a_i\}$. In general, there will be
degenerate sets of eigenvalues and the degeneracy arises
if one of the $a_i$ is zero. Using the triangular inequality it is easy
to show that
\begin{equation}
  2|a_i|\le|a_{i-1}|+|a_{i+1}|,
\end{equation}
i.e. $|a_i|$ are convex functions of discrete variables $i$. Then,
given the boundary condition in Eq. (\ref{a_1}), there could
at most be one $a_i=0$ in the solution for a given $\{\eta_i\}$,
which means that each eigenvalue set can be at most doubly degenerate.

Finding the number of  degeneracies of solutions is interesting but
difficult in general. We show that for $L=3$ (mod $4$) the number of 
such
degenerate sets of eigenvalues $\ge2^{(L+1)/2}$.

Proof: Consider the system of length $L=4m+3$, with $m$ being a non negative
integer. For any given set $\{\eta_i\}$, $i=1$ to $2m+2$, it is
possible to construct a solution $\{b_i\}$ of Eq. (\ref{eta}) with
$i\le2m+2$ which satisfies $b_0=1$ and $b_{2m+2}=0$.
Clearly, from  Eq.(\ref{eta}), if we have the solution $\{a_i\}$ 
corresponding to a particular set $\{\eta_j\}$, one can construct the 
solution $\{a_i'\}$ corresponding to $\{ \eta_j'= 
-\eta_j\}$  using 
$a_j' = (-1)^j a_j$. Using this symmetry we
extend $\{b_i\}$ ($i = 1 $ to $(L+1)/2$) to form a set $\{a_i\}$ for 
$i = 1$ to $L$  as follows:
\begin{eqnarray}
a_{i}&=&b_{i} \rm{~~~~~~~~~~~~~~~~~for~}i\le 2m+2, \\
     &=&(-1)^{i}b_{L+1-i} \rm{~~~for~}i>2m+2. 
\end{eqnarray}
This is a solution of Eq.(\ref{eta}) for the set $\{\eta'_i\}$ with 
\begin{eqnarray}
\eta_{i}'&=&\eta_{i} \rm{~~~~~~~~~~~~~~~~~for~} i\le 2m+2, \\
         &=&-\eta_{L+1-i} \rm{~~~~~~~~for~}i>2m+2.
\end{eqnarray}
and this solution $\{a_i\}$ satisfies the boundary conditions
$a_0=1$, $a_{L+1}=1$, and $a_{2m+2}=0$ (Fig.$1$). There are
$2^{2m+2}$ such solutions possible corresponding to all possible sets
of $\{\eta_i'\}$, and this gives the lower bound for the number of
degenerate solutions.

\begin{figure}
   \begin{center}
    \includegraphics[scale=0.25,angle=0]{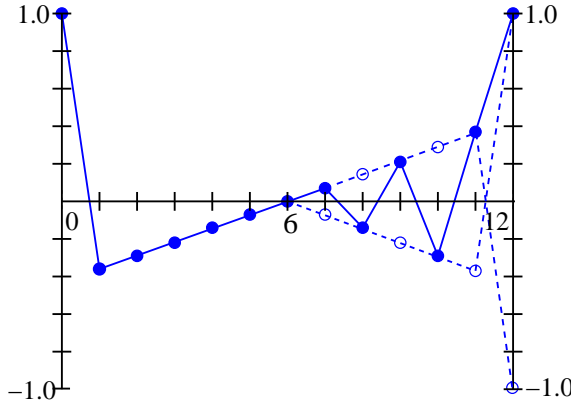}
    \caption{Schematic of a solution of Eq.(\ref{eta}) for $L=11$ with
$a_6=0$, and $a_0=a_{12}=1$. The filled circles represent the values
of the solution $\{a_i\}$.}
    \end{center}
    \label{figure1}
\end{figure}

A direct numerical calculation for $L\le20$ shows
that if $L\ne3$ (mod $4$), all $2^L$ sets of eigenvalues are
distinct. We present the degeneracies of the solutions in
Table.$1$. Calculation for simple choices of $\eta$ shows that
the degeneracies are possible only if $L=3$ (mod $4$).

For example,
consider $\eta_i=-1$ for $i=L$ and for the rest of the sites it is
$1$.  Then $a_i$ is of the form $a_i=1- \alpha i$, for all $i$. If we 
want this to be zero for $i=k$, we must have $\alpha = 1/k$.  Then, 
requiring that the equation (\ref{eta}) be satisfied at $i=L$, gives   
$3L=4k+1$, i.e. $L=3$ (mod $4$). Similarly for
the set with $\eta_{L-1}=-1$ and $1$ for rest of the sites imposes a
condition on length $3L=8k+1$, which is also a subset of $L=3$ (mod
$4$). Finding a general proof that degeneracies occur only if $L 
=3$ (mod $4$) remain an  open problem.

For each degenerate subspace there is a generalized eigenvector
linearly independent of the eigenvector corresponding to the
eigenvalue of the subspace. In general, let us denote them by
$|\{a_i\};n \rangle$, where $n=1$ for the eigenvector and $n=2$ for
the generalized eigenvector. For non-degenerate subspace $n$ can only
be $1$. The vectors satisfy the following equations
\begin{eqnarray}
  \mathbf{a}_i|\{a_j\};1\rangle &=& a_i|\{a_j\};1\rangle, \nonumber \\
  \mathbf{a}_i|\{a_j\};2\rangle &=& a_i|\{a_j\};2\rangle + \alpha_i
|\{a_j\};1\rangle,
  \label{gen}
\end{eqnarray}
where $\alpha$'s are complex numbers. Then using the Eq. (\ref{eta}) it can be shown
easily that $\alpha$'s satisfy the following equation
\begin{equation}
  \eta_i\alpha_i=\frac{1}{2}(\alpha_{i-1}+\alpha_{i+1}).
  \label{etaalpha}
\end{equation}
This is similar to the Eq. (\ref{eta}), except the boundary conditions
which are
\begin{equation}
  \alpha_0=\alpha_{L+1}=0.
\end{equation}
For a given set of $\{\eta_i\}$, these are $L$ simultaneous set of homogeneous 
linear equations which has infinitely many possible solutions. In
order to get a single solution we choose $\alpha_i=1$ if $a_i=0$, without
loss of generality. This corresponds to choosing a particular 
normalization of the rank $2$ eigenvectors.
The solution of both the equations (\ref{eta}) and
(\ref{etaalpha}) can be easily obtained numerically. 
The generalized eigenvectors and the Jordan block
form of the addition operators for the
system of size $L=3$ are given in the appendix.
%%%%%%%%%%%%%%%%%%%%%%%%%%%%%%%%%%%%%%%%%%%%%%%%%%%%%%%%%%%%%%%%%%
\begin{table}
  \begin{center}
    \begin{tabular}{|c||c||c|c|c|c|c|c|c|c|c|c|}
      \hline
      ~L~ & $g$ & $N_1$ & $N_2$ & $N_3$ & $N_4$ & $N_5$ & $N_6$ & $N_7$ &
$N_8$ & $N_9$ & $N_{10}$ \\
      \hline
      3 & 4 & 0 & 4 & & & & & & & &  \\
      \hline
      7 & 40 & 0 & 0 & 8 & 24 & & & & & & \\
      \hline
      11 & 136 & 0 & 0 & 0 & 8 & 0 & 120 & & & & \\
      \hline
      15 & 1304 & 0 & 0 &  0 & 4 & 32 & 48 & 288 & 560 & & \\
      \hline
      19 & 3024 & 0 & 0 & 0 & 0 & 0 & 8 & 0 & 288 & 0 & 2432 \\
      \hline
    \end{tabular}
    \caption{Degeneracies arise if one of the $a_i$ is
zero in a solution of
Eq.(\ref{eta}). In the table, $g$ denotes the total number of solutions with one of the $a_i=0$  i.e. the
total number of degenerate sets of solution. $N_i$ is the
number of solutions with the eigenvalue $a_i=0$.
Values for the other half of the system can be obtained using symmetry.}
  \end{center}
  \label{first}
\end{table}
%%%%%%%%%%%%%%%%%%%%%%%%%%%%%%%%%%%%%%%%%%%%%%%%%%%%%%%%%%%%%%%%%%%%%%%%%%%

\section{Matrix representation in the configuration basis}

Given the well defined action of the addition operators on the
generalized eigenvectors it is
possible to define a transformation
matrix $\mathbf{M}$ between the  configuration basis and the generalized 
eigenvector basis. 
\begin{equation}
|\{z_i\}\rangle = \sum_{j}\mathbf{M}_{\{z_i\},j}|\psi_j\rangle,
\end{equation}
where $|\{z_i\}\rangle$ is the  basis vector of $\Gamma$  corresponding to 
the height configuration
$\{z_i\}$ and $|\psi_j\rangle$ is the $j$th generalized eigenvector. Let us
express the configuration $|\{0\}\rangle$, with all sites empty, as a 
linear combination of all the generalized eigenvectors.
\begin{equation}
  |\{0\}\rangle = \sum_{j}c_j|\psi_j\rangle,
\end{equation}
where $c_j$s are constants. Then all the stable configurations can be
obtained by adding grains at properly chosen sites in $|\{0\}\rangle$.
\begin{equation}
  |\{z_i\}\rangle = \prod_i\mathbf{a}_i^{z_i}|\{0\}\rangle =
\sum_jc_j\prod_i\mathbf{a}_i^{z_i}|\psi_j\rangle,
\end{equation}
and hence
\begin{equation}
\mathbf{M}_{\{z_i\},j}=\langle \{z_i\}|\prod_i\mathbf{a}_i^{z_i}|\psi_j\rangle.
\end{equation}

The action of the addition operators on the generalized eigenvectors, 
for example Eq.(\ref{gen}) for model A, would generate the elements of 
the matrix $\mathbf{M}$. 
Given $\mathbf{M}$, we can get the eigenvectors of $\mathbf{a}_i$, in
the configuration basis, in 
particular, the steady state vector, by the inverse transformation 
\begin{equation}
|\psi_j\rangle = \mathbf{M}^{-1} |\{z_i\}\rangle.
\end{equation}
The  addition operators in the  
configuration basis  are obtained using the similarity transformation 
$\mathbf{M}\mathbf{a}^J_i\mathbf{M}^{-1}$. An explicit form of 
$\mathbf{M}$ for model A of length $L=3$ is given in the appendix.

\section{Determination of the steady state vector}
The time-evolution of the system is Markovian and the 
evolution operator $\mathbf{W}$ is defined by the master equation
\begin{equation}
  |P(t+1)\rangle=\mathbf{W}|P(t)\rangle,
\end{equation}
where $|P(t)\rangle$ and $|P(t+1)\rangle$ are the state of the
system at time $t$ and $t+1$, respectively. We can write the
time-evolution operator in terms of the addition operators as
\begin{equation}
  \mathbf{W}=\frac{1}{L}\sum_i\mathbf{a}_i.
\end{equation}
Then the common eigenvector of all the addition operators
corresponding to eigenvalue $1$ is the steady
state vector of the system. The steady state vector can be determined
in the stable configuration basis using the matrix $\mathbf{M}^{-1}$.
For model A of length $L=3$ the steady state vector is 
\begin{eqnarray}
  |S\rangle = &&\frac{13}{392}|0, 0, 0\rangle + \frac{1}{16} |1, 0, 0\rangle 
+ \frac{47}{392} |0, 1, 0\rangle + \frac{3}{16} |1, 1, 0\rangle \nonumber \\
   &&+  \frac{1}{16} |0, 0, 1\rangle + \frac{13}{98} |1, 0, 1\rangle 
+ \frac{3}{16} |0, 1, 1\rangle + \frac{3}{14} |1, 1, 1\rangle,
\end{eqnarray}
where the stable configurations are denoted by $|z_1,z_2,z_3\rangle$ with
$z_i$ as the height of the $i$th site.
The amplitude of each term in the expansion are the probability
of finding the steady state in the corresponding 
height configuration.

\begin{figure}
    \includegraphics[scale=0.75,angle=0]{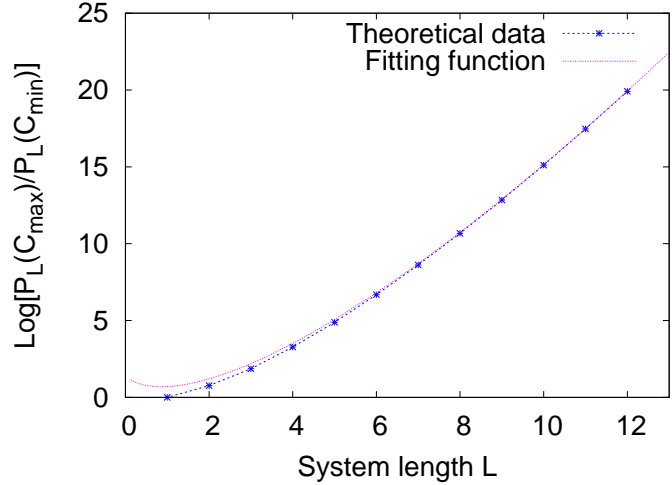}
    \caption{The ratio of the probability of the most probable
configuration $C_{max}$ (all occupied) and the least probable
configuration $C_{min}$ (all sites empty) plotted as a function of 
the system
length $L$. The fitting function $f(x)=a-bL+cL\log L$, with $a=1.50,
b=0.80$ and $c=0.94$.}
    \label{fig2}
\end{figure}

\section{Numerical Results}

Here we numerically calculate the exact steady state of model A for
different system length and discuss its properties.
As shown in Eq. (\ref{eta}) and Eq. (\ref{etaalpha}) the
eigenvalues $\{a_i\}$ and the off-diagonal matrix elements
$\{\alpha_i\}$ form sets of linear equations
for a given set of $\{\eta_i\}$. We solve them by LU
decomposition method. Because of the tridiagonal structure
of the equations, only $\mathcal{O}(2^L)$ number of steps are required
to get the solution. The maximum number of steps ($\mathcal{O}(2^{3L})$)
are required for the inversion of the transformation matrix
$\mathbf{M}$. We have used the Gauss-Jordan elimination method
for the inversion. It is important to note that, the
maximum system length $L$, possible to treat by this method,
is determined by the limited memory size of the computers,
and not by the computation time. Using desktop computers we were able
to determine $\mathbf{M}$ exactly for systems of size $L\le12$. 

We note that as $L$ is increased, the second largest eigenvalue of
$\mathbf{W}$
tends to $1/2$. Thus, the gap between the largest and the next largest
eigenvalue of the relaxation matrix does not tend to zero. 
This gap measures the relaxation time of the system in terms of the
macro-time unit of interval between addition of grains. However, the
average duration of an avalanche measured in terms of micro-time unit
of duration of a single toppling event does diverge, as system size
increases.

An interesting question is the extent of  variation between 
probabilities of
different configurations in the steady state. In the one-dimensional
Oslo model, for a system of $L$ sites, the ratio of probabilities of
the most probable to the least probable configuration varies as
$\exp(L^3)$\cite{oslo}.
However in model A, we find that the ratio is not quit as large, and it
only varies approximately as $\exp(0.94L\log L)$ (Fig.\ref{fig2}) for 
large $L$. 

\begin{figure}
    \includegraphics[scale=0.75,angle=0]{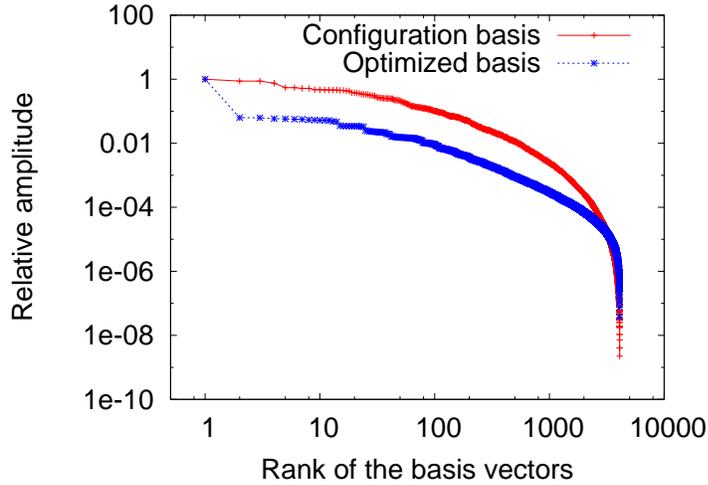}
    \caption{The amplitudes, normalized with its largest value, corresponding to the basis vectors
 in the steady state plotted
as a function of the rank of the basis vectors.
The vectors are arranged in decreasing orders of their
amplitudes. The plot is given for the configuration basis and the
optimized basis for model A of size $L=12$.}
    \label{fig3}
\end{figure}

This suggests that possibly the
exact steady state has a product measure. To check this we define a 
product basis $|\psi'\rangle =
\displaystyle\prod_i|\psi_i'\rangle$,
where $|\psi_i'\rangle$ could be any one of the two orthogonal vectors
\begin{eqnarray}
  |1'\rangle &=& \cos\phi_i|1\rangle + \sin\phi_i|0\rangle, \nonumber \\
  |0'\rangle &=& \sin\phi_i|1\rangle - \cos\phi_i|0\rangle,
\end{eqnarray}
with $\phi_i$ a real number. Then in this basis the steady state can
be written as
\begin{equation}
  |S\rangle = \sum_{\psi'} P(\psi')|\psi'\rangle.
\end{equation}
We choose $\{\phi_i\}$ so that the ratio between the amplitudes
of basis vectors with next-largest and largest 
amplitudes becomes as small as possible (this would become zero, 
if the state was a product measure state). 
In Fig.\ref{fig3}, we have plotted for system of 
size $L=12$, the relative amplitudes in both configuration
basis and the optimized product basis
as a function of the rank of the basis vectors with the vectors
arranged in decreasing orders of their amplitudes. In the
optimized basis the second highest probability is only $10$ times
smaller than the highest probability. This implies that the steady
state measure is not a product measure.

\begin{figure}
    \includegraphics[scale=0.75,angle=0]{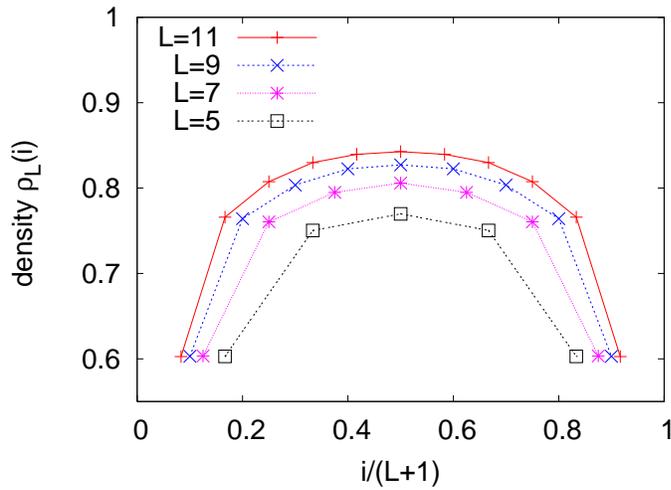}
    \caption{Average steady state density $\rho_{L}(i)$ at site $i$ for
model A of different length $L$.}
    \label{fig4}
\end{figure}
\begin{table}
  \begin{center}
    \begin{tabular}{|c|c|c|}
      \hline
      $ ~ ~ ~ ~ ~ L ~ ~ ~ ~ ~ $ & $ ~ ~ ~ ~ ~ ~ \bar{\rho}_L ~ ~ ~ ~ ~ ~ $ & $ ~ ~ ~ ~ ~ ~ \rho_L(x_m) ~ ~ ~ ~ ~ ~ $ \\
      \hline
      2 & 0.583333 & 0.583333 \\
      \hline
      3 & 0.634354 & 0.709184 \\
      \hline
      4 & 0.669262 & 0.737000 \\
      \hline
      5 & 0.695210 & 0.769704 \\
      \hline
      6 & 0.715472 & 0.786491 \\
      \hline
      7 & 0.731879 & 0.805897 \\
      \hline
      8 & 0.745514 & 0.816009 \\
      \hline
      9 & 0.757080 & 0.827217 \\
      \hline
      10 & 0.767051 & 0.834600 \\
      \hline
      11 & 0.775760 & 0.842665 \\
      \hline
      12 & 0.783451 & 0.848054 \\
      \hline
    \end{tabular}
    \caption{The values of particle density in the steady state 
    for the model A of different length $L$. Here $\bar{\rho}_L$ denotes the
    steady state particle density averaged over all sites and $\rho_L(x_m)$ denotes the
    steady state particle density at the central site.}
  \end{center}
  \label{second}
\end{table}
\begin{table}
  \begin{center}
    \begin{tabular}{|c|c|c|c|}
      \hline
      ~L~ & $1/\bar{\rho}_\infty$ & $B$ & $\nu_\perp$ \\
      \hline
      \hline
      3 & ~ ~ 1.061 ~ ~ & ~ ~ 1.128 ~ ~ & ~ ~ 0.656 ~ ~ \\
      \hline
      4 & 1.049 & 1.132 & 0.641 \\
      \hline
      5 & 1.053 & 1.132 & 0.646 \\
      \hline
      6 & 1.049 & 1.131 & 0.640 \\
      \hline
      7 & 1.050 & 1.131 & 0.641 \\
      \hline
      8 & 1.049 & 1.130 & 0.639 \\
      \hline
      9 & 1.049 & 1.130 & 0.639 \\
      \hline
      10 & 1.049 & 1.130 & 0.639 \\
      \hline
      11 & 1.049 & 1.130 & 0.639 \\
      \hline
    \end{tabular}
    \caption{The sequential fit of the functional form in Eq. (\ref{rhobar}) to the
    data for average particle density for model A of different length $L$ given in Table $2$.}
  \end{center}
  \label{third}
\end{table}

The steady state density for different sites
are plotted in Fig. \ref{fig4} for different system sizes.
Amongst the different fitting forms that we tried,  the following 
functional form gives the best fit
\begin{equation}
  \frac{1}{\rho_L(x)} = \frac{1}{\bar{\rho}_\infty} + b[\frac{1}{(x+d)^{\nu_\perp}} + \frac{1}{(L+1-x+d)^{\nu_\perp}}],
\end{equation}
where $\bar{\rho}_\infty, b, \nu_\perp$ and $d$ are real numbers. 
Using this functional form the steady state particle density
averaged over all sites for system of size $L$ can be written as
\begin{equation}
  \frac{1}{\bar{\rho}_L} = \frac{1}{\bar{\rho}_\infty} + \frac{B}{(L+\delta)^{\nu_\perp}},
  \label{rhobar}
\end{equation}
where $B$ is a real number and $\bar{\rho}_\infty$
is the asymptotic value of the average particle density.
The exact value of $\bar{\rho}_L$ and the particle density
at the central site $\rho_L(x_m)$ are listed in the Table $2$
for different system sizes. 
The sequential fitting method is used to find the values of 
$\bar{\rho}_\infty$, $B$, $\nu_\perp$ and $\delta$ from these data. For a given choice
of $\delta$, these values are obtained numerically by solving the Eq. (\ref{rhobar})
for three consecutive lengths $L-1$, $L$ and $L+1$. Best
convergence of the values of $\bar{\rho}_\infty$, $B$ and $\nu_\perp$ are obtained for
$\delta=1.1$, which are tabulated in Table $3$.
The asymptotic value of the average particle density
converges to $\bar{\rho}_\infty =0.953$ which is close
to the more precise estimate $0.94885(7)$, from Monte 
Carlo simulations  \cite{dickman1}.

\section{Concluding remarks.}
For a general SASM with $N$ sites, the calculation of eigenvalues 
involves solving $N$ coupled polynomial equations in $N$ variables. This 
can be done in polynomial time in $\Omega$, the number of stable 
configurations of the model. These are then used to construct the 
transformation matrix ${\bf M}$ of size $\Omega \times \Omega$. Finally 
inverting the matrix ${\bf M}$ gives us the eigenvectors of the 
evolution operator, in particular the steady state.

Of course, to determine the steady state of any Markov chain on $\Omega$ 
states, we need to determine the eigenvectors of the evolution matrix of 
size $\Omega \times \Omega$. The point here is that the specification of 
the toppling rules does not directly specify the evolution matrix, and 
determining the matrix elements of the latter from the toppling rules is 
computationally very nontrivial. Using  the 
abelian property, we are able to tackle this problem.

For a generic model with some parameters, e.g. the model B, except for 
special symmetries, one does not expect degeneracies in eigenvalues to
occur 
for a generic value of the parameters.  For special values of the 
parameters, if there is a  non-trivial Jordan 
block structure of the evolution operator, it would  show up in the 
time-dependent correlation
functions of the model by the presence of terms of the type $ t \exp( 
-\lambda_j 
t)$, in addition to the usual sum of terms of the type $\exp( 
-\lambda_j 
t)$.

In particular we have
explicitly calculated the steady state for a specific model (model A
in section 2) of length L$\le12$. Extrapolating the results
we determined the asymptotic density profile in the steady state.
The power-law profile of deviations from the mean value near the ends
would be important for determining the avalanche exponents of the model 
\cite{lubeck2}. This remains an interesting open problem.

\begin{acknowledgement}
The work of DD is supported in part by a J. C. Bose Fellowship of the 
Government of India.
\end{acknowledgement}

\appendix
\section{Appendix}

Here we give some details of the explicit calculation of the steady 
state, and the matrix representation of addition operators for model A 
of length $L=3$. 

The eight sets of eigenvalues  obtained by solving Eq.(\ref{eta})  are 
$(1, 1, 1)$, $(-1, 1, -1)$, $(\frac{1}{3}, -\frac{1}{3},\frac{1}{3})$, 
$(-\frac{1}{3}, -\frac{1}{3}, -\frac{1}{3})$, $ 
(\frac{1}{2}, 0, -\frac{1}{2})$, and $(-\frac{1}{2}, 0, \frac{1}{2})$ 
with the last two sets repeated twice.

For writing the matrix structure of the addition operators, we choose 
the  order of the eigenvectors same  
as the order of the eigenvalues mentioned above. For the degenerate subspace we 
order the eigenvector $|\{a_i\};1\rangle$, defined in 
Eq.(\ref{gen}), before the generalized eigenvector $|\{a_i\};2\rangle$.  Then in 
this basis the matrices corresponding to the addition operators have the 
following Jordan block form. \begin{align}
 \textbf{a}^J_1 = 
  \begin{pmatrix}
    1 ~&~ 0 ~&~ 0 ~&~ 0 ~&~ 0 ~&~ 0 ~&~ 0 ~&~ 0 \\
    0 ~& -1 &~ 0 ~&~ 0 ~&~ 0 ~&~ 0 ~&~ 0 ~&~ 0 \\
    0 ~&~ 0 ~&~ \frac{1}{3} ~&~ 0 ~&~ 0 ~&~ 0 ~&~ 0 ~&~ 0 \\
    0 ~&~ 0 ~&~ 0 ~& -\frac{1}{3} &~ 0 ~&~ 0 ~&~ 0 ~&~ 0 \\
    0 ~&~ 0 ~&~ 0 ~&~ 0 ~&~ \frac{1}{2} ~&~ \frac{1}{2} ~&~ 0 ~&~ 0 \\
    0 ~&~ 0 ~&~ 0 ~&~ 0 ~&~ 0 ~&~ \frac{1}{2} ~&~ 0 ~&~ 0 \\
    0 ~&~ 0 ~&~ 0 ~&~ 0 ~&~ 0 ~&~ 0 ~& -\frac{1}{2} & -\frac{1}{2} \\
    0 ~&~ 0 ~&~ 0 ~&~ 0 ~&~ 0 ~&~ 0 ~&~ 0 ~& -\frac{1}{2} 
 \end{pmatrix},
\end{align}
\begin{align}
 \textbf{a}^J_2 = 
  \begin{pmatrix}
    1 ~&~ 0 ~&~ 0 ~&~ 0 ~&~ 0 ~&~ 0 ~&~ 0 ~&~ 0 \\
    0 ~&~ 1 ~&~ 0 ~&~ 0 ~&~ 0 ~&~ 0 ~&~ 0 ~&~ 0 \\
    0 ~&~ 0 ~& -\frac{1}{3} &~ 0 ~&~ 0 ~&~ 0 ~&~ 0 ~&~ 0 \\
    0 ~&~ 0 ~&~ 0 ~& -\frac{1}{3} &~ 0 ~&~ 0 ~&~ 0 ~&~ 0 \\
    0 ~&~ 0 ~&~ 0 ~&~ 0 ~&~ 0 ~&~ 1 ~&~ 0 ~&~ 0 \\
    0 ~&~ 0 ~&~ 0 ~&~ 0 ~&~ 0 ~&~ 0 ~&~ 0 ~&~ 0 \\
    0 ~&~ 0 ~&~ 0 ~&~ 0 ~&~ 0 ~&~ 0 ~&~ 0 ~&~ 1 \\
    0 ~&~ 0 ~&~ 0 ~&~ 0 ~&~ 0 ~&~ 0 ~&~ 0 ~&~ 0 
 \end{pmatrix},
\end{align}
\begin{align}
 \textbf{a}^J_3 = 
  \begin{pmatrix}
    1 ~&~ 0 ~&~ 0 ~&~ 0 ~&~ 0 ~&~ 0 ~&~ 0 ~&~ 0 \\
    0 ~& -1 &~ 0 ~&~ 0 ~&~ 0 ~&~ 0 ~&~ 0 ~&~ 0 \\
    0 ~&~ 0 ~&~ \frac{1}{3} ~&~ 0 ~&~ 0 ~&~ 0 ~&~ 0 ~&~ 0 \\
    0 ~&~ 0 ~&~ 0 ~& -\frac{1}{3} &~ 0 ~&~ 0 ~&~ 0 ~&~ 0 \\
    0 ~&~ 0 ~&~ 0 ~&~ 0 ~& -\frac{1}{2} & -\frac{1}{2} &~ 0 ~&~ 0 \\
    0 ~&~ 0 ~&~ 0 ~&~ 0 ~&~ 0 ~& -\frac{1}{2} &~ 0 ~&~ 0 \\
    0 ~&~ 0 ~&~ 0 ~&~ 0 ~&~ 0 ~&~ 0 ~&~ \frac{1}{2} ~&~ \frac{1}{2} \\
    0 ~&~ 0 ~&~ 0 ~&~ 0 ~&~ 0 ~&~ 0 ~&~ 0 ~&~ \frac{1}{2} 
  \end{pmatrix},
\end{align}
The transformation matrix $\mathbf{M}$, discussed in section $5$, between
the generalized eigenvector basis and the
configuration basis has the following form
\begin{align}
  \textbf{M} = 
  \begin{pmatrix}
    1 ~&~ 1 ~&~ 1 ~&~ 1 ~&~ 1 ~&~ 1 ~&~ 1 ~&~ 1 \\
    1 ~& -1 ~&~ 1/3 ~& -1/3 ~&~ 1 ~&~ 1/2 ~& -1 ~& -1/2 \\
    1 ~&~ 1 ~& -1/3 ~& -1/3 ~&~ 1 ~&~ 0 ~&~ 1 ~&~ 0 \\
    1 ~& -1 ~& -1/9 ~&~ 1/9 ~&~ 1/2 ~&~ 0 ~& -1/2 ~&~ 0 \\
    1 ~& -1 ~&~ 1/3 ~& -1/3 ~& -1 ~& -1/2 ~&~ 1 ~&~ 1/2 \\
    1 ~&~ 1 ~&~ 1/9 ~&~ 1/9 ~& -3/4 ~& -1/4 ~& -3/4 ~& -1/4 \\
    1 ~& -1 ~& -1/9 ~&~ 1/9 ~& -1/2 ~&~ 0 ~&~ 1/2 ~&~ 0 \\
    1 ~&~ 1 ~& -1/27 ~& -1/27 ~& -1/4 ~&~ 0 ~& -1/4 ~&~ 0 
  \end{pmatrix},
\end{align}
where the configuration basis vectors are chosen in the following order
$(0, 0, 0)$, $(1, 0, 0)$, $(0, 1, 0)$, $(1, 1, 0)$, $(0, 0, 1)$,
$(1, 0, 1)$, $(0, 1, 1)$, and $(1, 1, 1)$. The matrix is non-singular,
and the inverse can be calculated numerically.
Using the similarity transformation $\mathbf{M}\mathbf{a}_1^J\mathbf{M}^{-1}$ we find matrix
representation of the addition operator $\mathbf{a}_1$ in the configuration
basis.
\begin{align}
  \textbf{a}_1 =
  \begin{pmatrix}
    0 ~&~ \frac{2}{7} ~&~ 0 ~&~ \frac{4}{49} ~&~ 0 ~&~ 0 ~&~ 0 ~&~ 0 \\
    1 ~&~ 0 ~&~ 0 ~&~ 0 ~&~ 0 ~&~ \frac{1}{24} ~&~ 0 ~&~ \frac{1}{9} \\
    0 ~&~ \frac{4}{7} ~&~ 0 ~&~ \frac{22}{49} ~&~ 0 ~&~ 0 ~&~ 0 ~&~ 0 \\
    0 ~&~ 0 ~&~ 1 ~&~ 0 ~&~ 0 ~&~ \frac{1}{12} ~&~ 0 ~&~ \frac{19}{72} \\
    0 ~&~ 0 ~&~ 0 ~&~ 0 ~&~ 0 ~&~ \frac{7}{24} ~&~ 0 ~&~ \frac{1}{9} \\
    0 ~&~ \frac{1}{7} ~&~ 0 ~&~ \frac{16}{49} ~&~ 1 ~&~ 0 ~&~ 0 ~&~ 0 \\
    0 ~&~ 0 ~&~ 0 ~&~ 0 ~&~ 0 ~&~ \frac{7}{12} ~&~ 0 ~&~ \frac{37}{72} \\
    0 ~&~ 0 ~&~ 0 ~&~ \frac{1}{7} ~&~ 0 ~&~ 0 ~&~ 1 ~&~ 0 
  \end{pmatrix},
\end{align}
The other operators can also be determined similarly.


\begin{thebibliography}{}

\bibitem{deepak} Dhar D.: Theoretical studies of self-organized
criticality. Physica A \textbf{369}, 29 (2006) 

\bibitem{manna} Manna S. S.: Two-state model of self-organized
criticality. J. Phys. A: Math. Gen. \textbf{24}, L363 (1991)

\bibitem{rice} Frette V., Christensen K., Mathe-Sorensen A., Feder J., 
Jossang T. and Meakin P.: Avalanche dynamics in a pile of rice. Nature 
\textbf{379}, 49 (1996)

\bibitem{chessa} Chessa A., Stanley H. E., Vespignani A., and Zapperi 
S.: Universality in sandpiles. Phys. Rev E. \textbf{59}, R12 (1999)

\bibitem{campelo} Dickman R., and Campelo J. M. M.: Avalanche exponents 
and corrections to scaling for a stochastic sandpile. Phys. Rev. E 
\textbf{67}, 066111 (2003)

\bibitem{btw} Bak P., Tang C., and Wiesenfeld K.: Self-organized 
criticality: An explanation of the 1/f noise. Phys. Rev. Lett. 
\textbf{59}, 381 (1987); Self-organized criticality. Phys. Rev. A 
\textbf{38}, 364 (1988)

\bibitem{asap} Povolotsky A. M., Priezzhev V. B., and Hu C. K.: The 
asymmetric avalanche process. J. Stat Phys \textbf{3}, 1149 (2003)

\bibitem{rittenberg} Alcaraz F. C., and Rittenberg V.: Directed abelian 
algebras and their applications to stochastic models. arXiv:0806.1303

\bibitem{oslo} Dhar D.: Steady state and relaxation spectrum of the Oslo 
rice-pile model. Physica \textbf{A} \textbf{340}, 535 (2004)

\bibitem{kloster} Kloster M., Maslov S., Tang C.: Exact solution of a 
stochastic directed sandpile model. Phys. Rev. E \textbf{63}, 026111 
(2001)

\bibitem{paczuski} Paczuski M., Bassler K. E.: Theoretical
results for sandpile models of self-organized criticality with
multiple topplings. Phys. Rev. E \textbf{62}, 5347 (2000)

\bibitem{dickman1} Dickman R., Alva M., Mu$\tilde{n}$oz, Peltola J., 
Vespignani A., Zapperi S.: Critical behavior of a one-dimensional
fixed-energy stochastic sandpile. Phys. Rev. {\bf E 64}, 056104 (2001)

\bibitem{dickman2} Stilck J. F., Dickman R. and Vidigal R. R.: Series
expansion for a stochastic sandpile. J. Phys. A: Math. Gen.
{\bf  37}, 1145 (2004)

\bibitem{dickman3} Vidigal R. R. and Dickman R.: Asymptotic behavior
of the order parameter in a stochastic sandpile. J. Stat. Phys.
\textbf{118}, 1 (2005)

\bibitem{pietronero} Pietronero L., Vespignani A., and Zapperi S.:
Renormalization scheme for self-organized criticality in sandpile
models. Phys. Rev. Lett. \textbf{72}, 1690 (1994); Vespignani A., 
Zapperi S., and Pietronero L.: Phys. Rev. E \textbf{57}, 1711 (1995)

\bibitem{biham} Ben-Hur A., and Biham O.: Universality in sandpile
models. Phys. Rev. E \textbf{53}, R1317 (1996); Milshtein E., Biham
O., and Solomon S.: Universality classes in isotropic, Abelian, and
non-Abelian sandpile models. Phys. Rev. E \textbf{58}, 303 (1998)

\bibitem{lubeck} Lubeck S.: Moment analysis of the probability
distribution of different sandpile models. Phys. Rev. E \textbf{61}, 204 (2000)

\bibitem{menech} Menech M. D.: Comment on ``Universality in
sandpiles''. 
Phys. Rev. E \textbf{70}, 028101 (2004)

\bibitem{mohanty1} Mohanty P. K.  and  Dhar D.: Generic sandpiles have 
directed percolation exponents, Phys. Rev. Lett. 
\textbf{89}, 104303 (2002)

\bibitem{bonachela} Bonachela J. A. Ramasco J. J. Chate H., Dornic I. 
and Munoz M. A., Sticky grains do not change the universality class of 
isotropic sandpiles,  Phys. Rev. E \textbf{74}, 050102 (2006)

\bibitem{mohanty2} Mohanty P. K.  and Dhar D.: Critical behavior of 
sandpile models with sticky grains, Physica A \textbf{384}, 34 (2007)

\bibitem{sasm} Dhar D.: Some results and a conjecture for Manna's
stochastic sandpile model. Physica A \textbf{270}, 69 (1999)

\bibitem{lubeck2} Lubeck S. and Dhar D.: Continuously varying exponents 
in sandpile models. J. Stat. Phys. {\bf 102}, 1 (2001)

\end{thebibliography}
\end{document}